\documentclass{PoS}
\usepackage{natbib}

\title{Local stars formed at z>10: a sample extracted from the SDSS}

\ShortTitle{Local stars formed at z>10: a sample extracted from the SDSS}

\author{\speaker{Luca Sbordone}\\
        Max-Planck Institut f\"ur Astrophysik - Garching, Germany\\
        GEPI, Observatoire de Paris, CNRS, Universit\'e Paris Diderot, France}

\author{Piercarlo Bonifacio\\
        GEPI, Observatoire de Paris, CNRS, Universit\'e Paris Diderot, France\\
        INAF - Osservatorio Astronomico di Trieste, Italy}
        
\author{Elisabetta Caffau\\
		ZAH -- Landessternwarte Heidelberg, Germany,\\
       GEPI, Observatoire de Paris, CNRS, Universit\'e Paris Diderot, France}

\author{Hans-G\"unter Ludwig\\
		ZAH -- Landessternwarte Heidelberg, Germany,\\
       GEPI, Observatoire de Paris, CNRS, Universit\'e Paris Diderot, France}

\abstract{As the Universe emerged from its initial hot and dense phase, its chemical composition was extremely simple, being limited to stable H and He isotopes, and traces of Li. The first stars that formed had such initial composition. However, they quickly began to produce a whole array of heavier nuclei, polluting the interstellar medium. While none among these first stars has been detected to date, an increasing sample exists of their direct descendants, stars with a fraction of heavy elements of the order of 1/1000 of the solar value, or less. In most cases, such stars should have formed at redshift of about 10 or beyond, and their chemical composition can provide crucial constraints to the nature of the very first stars. 
Extremely metal poor (EMP) stars are exceedingly rare. We used the low resolution spectra obtained by the Sloan Digital Sky Survey (SDSS) to search for EMP candidates: results of VLT-UVES high resolution follow-up for 16 of them is presented here. A newly developed automatic abundance analysis and parameter determination code, MyGIsFOS, has been employed to analyze the detailed chemical abundances of such stars.}

\FullConference{11th Symposium on Nuclei in the Cosmos\\
		 19-23 July 2010 \\
		 Heidelberg, Germany.}

\begin{document}

\section{Target selection and low-resolution metallicity estimate}

Candidate EMP stars were searched among stars with colors compatible with Halo the F-G turn-ff population, and having been observed spectroscopically by SDSS (up to data release 6). SDSS low-resolution (R=2000) spectra were analyzed by means of an automatic abundance-estimation code, comparing a number of metallic features against a grid of synthetic spectra. The stellar effective temperature was estimated by means of a theoretical calibration of the T$_{eff}$ / (g-z) relationship. The gravity was kept fixed at $\log g = 4$ in this candidate-selection phase. See \cite{ludwig08} for a more detailed description of the candidate selection phase.

\section{Automated abundance analysis with MyGIsFOS}

The candidate EMP stars selected were then observed by means of UVES@VLT at a typical resolution of about R=30,000,  S/N$\sim$40-50 per pixel. Three among them appeared to be carbon-enhanced metal-poor stars and their full analysis is presented by \cite{behara10}. Sixteen more objects, not showing anomalous CNO abundances, are presented in this contribution.

\begin{figure}
\begin{center}
\includegraphics[width=.9\textwidth]{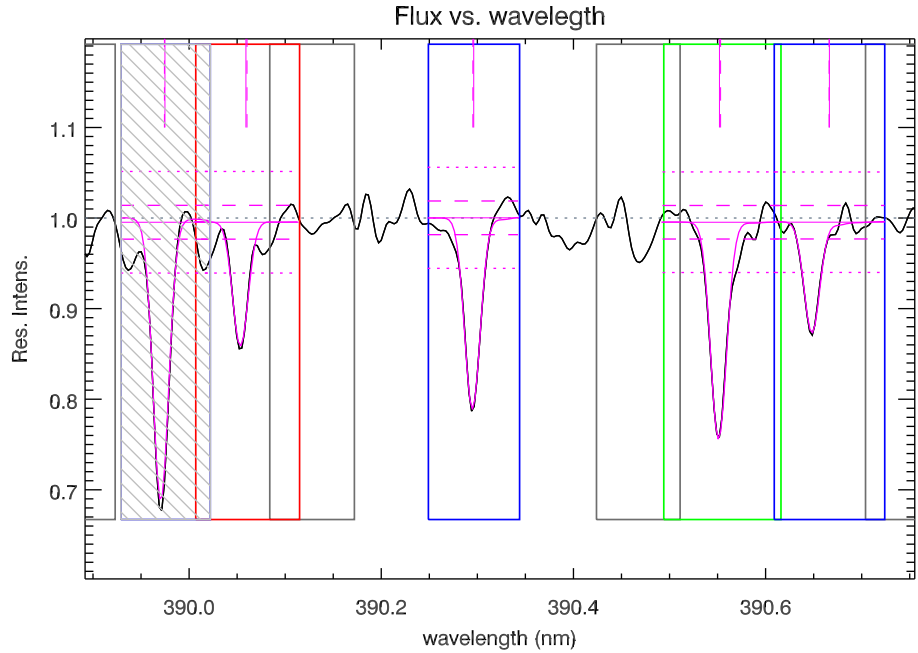} 
\caption{An example of a set of MyGIsFOS final fits over Fe I (blue boxes), Si I (green box), and Sc II (red box) features. Dark grey boxes outline pseudo-continuum regions. The light grey, shaded box is a Fe I feature which has been rejected as bad.} 
\label{mygisfos} 
\end{center}
\end{figure}

A new, fully automatic parameter estimation and abundance analysis code has been developed and applied to the full chemical analysis of these stars. The MyGIsFOS code works by replicating a standard manual abundance analysis where T$_{eff}$ is provided from outside (e.g. from photometry calibration), gravity, microturbulence and [Fe/H] are derived by measuring selected Fe I and Fe II features, and, after the atmosphere parameters are established, a full chemical analysis can be performed by measuring lines of all the other significant elements. The user selects a number of spectral regions for pseudo-continuum estimation, as well as all the needed metallic lines. Each line is then fitted against a grid of pre-calculated synthetic spectra varying in T$_{eff}$, $\log g$, microturbulence, [Fe/H], and [$\alpha$/Fe], by a $\chi^2$ minimization scheme. Stellar parameters are iterated from first-guess values to the final ones. An example of MyGIsFOS output is shown in Fig. \ref{mygisfos}.

Parameter estimation and full chemical analysis with MyGIsFOS takes beween 15 and 30 seconds per star on a mainstream PC. The code is currently available on collaboration basis, and is foreseen to be released to the community after debugging, performance assessment, and documentation are completed.


\section{Abundance analysis results}

\begin{figure}
\begin{center}
\includegraphics[width=.6\textwidth]{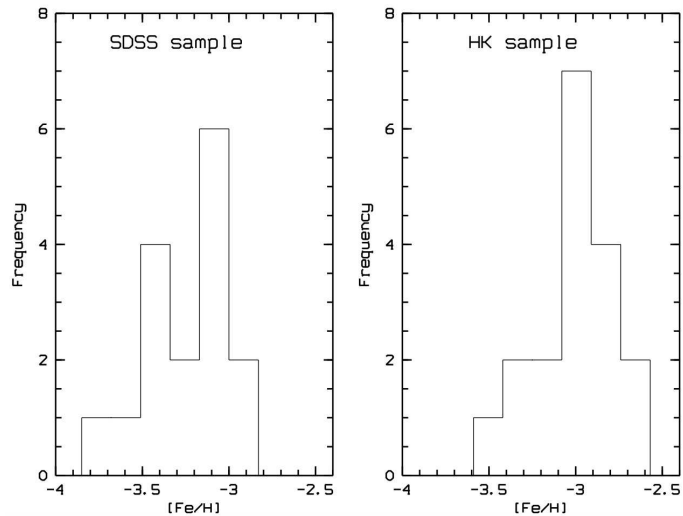} 
\caption{Comparison of the metallicity distribution of the sample of EMP stars extracted from SDSS (left panel) with the one presented in the ESO large program ``First Stars'' (right panel)} 
\label{methisto} 
\end{center}
\end{figure}

Compared to the sample analyzed in the large program ``First Stars'' \citep[][and references therein]{bonifacio09} the new sample appears remarkably more metal poor, assessing the effectiveness of using SDSS for target selection (see Fig. \ref{methisto}). The selection strategy proved highly effective: all the selected targets were confirmed as EMP by the high resolution analysis.

\begin{figure}
\begin{center}
\includegraphics[width=.9\textwidth]{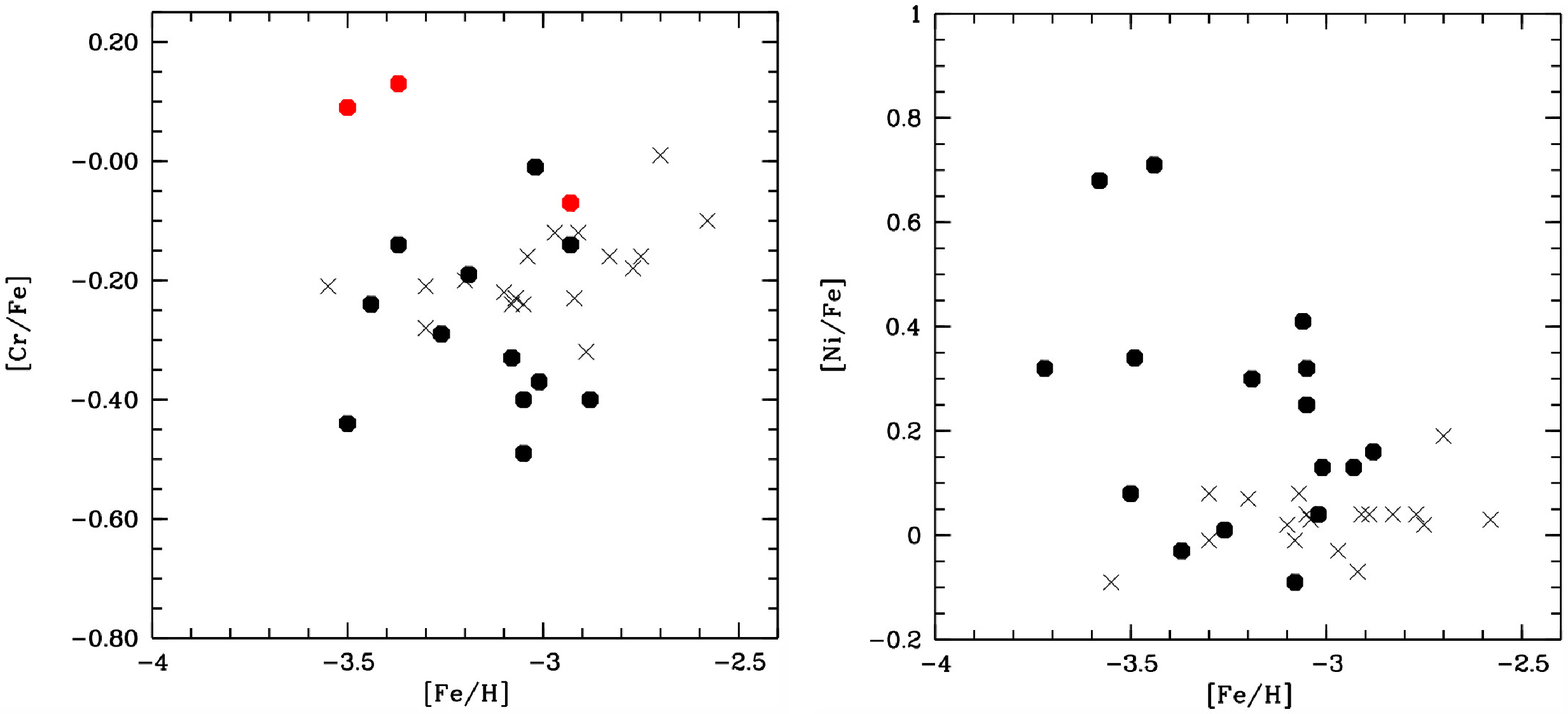} 
\caption{In the left panel, [CrI/Fe] is plotted against [Fe/H] for the SDSS sample (filled dots) and the ``First Stars'' sample (crosses). The three red filled dots correspond to Cr II measurements in three stars of the SDSS sample. In the right panel, [Ni/Fe] is plotted against [Fe/H], with the same symbols.} 
\label{crandni} 
\end{center}
\end{figure}

In fig. \ref{crandni} we present [Cr/Fe] and [Ni/Fe] plotted against [Fe/H] for the new sample along with the ``First Stars'' sample. In the left panel the three [CrII/Fe] measurements appear discrepant with respect of the [CrI/Fe] measurements for the same stars, underlining the importance of NLTE effects for Cr in EMP stars \citep{bonifacio09,bergemann10} which were not accounted for in the present analysis.

Of particular interest is the evidence of a sizable population of Ni-rich stars at extremely low metallicities: 8 stars out of 16 show [Ni/Fe]$>$0.2. While the current analysis is preliminary, it would appear that this result is indeed robust. A final analysis, including temperatures based on 3D-hydrodynamical H$\alpha$ profiles, is currently underway.

\begin{thebibliography}{99}

\bibitem[Behara et 
al.(2010)]{behara10} Behara, N.~T., Bonifacio, P., Ludwig, H.-G., Sbordone, L., Gonz{\'a}lez Hern{\'a}ndez, J.~I., \& Caffau, E.\ 2010, A\&A, 513, A72 

\bibitem[Bergemann 
\& Cescutti(2010)]{bergemann10} Bergemann, M., \& Cescutti, G.\ 2010, arXiv:1006.0243 

\bibitem[Bonifacio et 
al.(2009)]{bonifacio09} Bonifacio, P., et al.\ 2009, A\&A, 501, 519 

\bibitem[Ludwig et al.(2008)]{ludwig08} Ludwig, H.-G., 
Bonifacio, P., Caffau, E., Behara, N.~T., Gonz{\'a}lez Hern{\'a}ndez, 
J.~I., \& Sbordone, L.\ 2008, Physica Scripta Volume T, 133, 014037 

\end{thebibliography}
\end{document}